# Intrinsic robust antiferromagnetism at manganite surfaces and interfaces

S. Valencia[1]*, L. Peña[2], Z. Konstantinovic[2], Ll. Balcells[2], R. Galceran[2], D. Schmitz[1], F. Sandiumenge[2], M. Casanove[3], and B. Martínez[2]

[1] Helmholtz-Zentrum-Berlin. Albert-Einstein-Str. 15, 12489 Berlin (Germany)

[2] Institut de Ciéncia de Materials de Barcelona-CSIC. Campus de la UAB, 08193 Bellaterra (Spain) [3] Centre d´Elaboration de Matériaux et d´Etudes Structurales (CNRS-CEMES). BP 94347, 29 rue Jeanne Maarving, 31055 Toulouse (France)

Abstract:

Ferromagnetic/metallic manganese perovskites, such as $La_{2/3}Sr_{1/3}MnO_3$ (LSMO) are promising materials for the design and implementation of novel spintronic devices working at room temperature. However, their implementation in practical applications has been severely hampered due to the breakdown of their magnetotransport properties at temperatures well below their magnetic transition temperature. This breakdown has been usually associated to surface and interface related problems but its physical origin has not been clearly established yet. In this work we investigate the interface between $La_{2/3}Sr_{1/3}MnO_3$ (LSMO) thin films and different capping layers by means of x-ray linear dichroism and transport measurements. Our data reveal that, irrespective to the capping material, LSMO/capping layer bilayers exhibit an antiferromegnetic/insulating phase at the interface, likely to originate from a preferential occupancy of Mn $3d$ $3z^2-r^2$ $e_g$ orbitals. This phase, which extends ca. 2 unit cells, is also observed in an uncapped LSMO reference sample thus, pointing to an intrinsic interfacial phase separation phenomenon, likely to be promoted by the structural disruption and symmetry breaking at the LSMO free surface/interface. These experimental observations strongly suggest that the structural disruption at the LSMO interfaces play a major role on the observed depressed magnetotransport properties in manganite-based magnetic tunneling junctions and it is at the origin of the so-called dead layer.

Complex oxides have revealed to be one of the most interesting classes of materials due to their amazing variety of properties of strong theoretical and technological interest, including superconductivity, ferromagnetism, ferroelectricity, etc. In particular, transition metal oxides are especially relevant since they present large electronic correlations leading to a strong competition between various degrees of freedom[1]. Among them, manganites offer a plethora of possibilities both from a fundamental and applied physics point of view. Manganites are a complex system exhibiting a broad range of physical phenomena, including large spin polarization and colossal magnetoresistance[2]. These properties make them very appealing for the development of novel concepts for the implementation of oxide-based spintronic devices. Especially relevant is the case of $La_{0.67}Sr_{0.33}MnO_3$ (LSMO) which exhibits the highest magnetic transition temperature ($T_C \sim 370K$) of this family of materials[3], and therefore holds promise to be implemented in room temperature working magnetic tunneling devices. However, state of the art LSMO-based junctions show vanishing small tunneling magnetoresistance (TMR) values above 280 K (Ref. 4). The reason for this is usually attributed to the existence of a so-called dead layer, of a few unit cells width, at the interface between the manganite and the tunnel barrier whose origin is not yet well understood. TMR is a spin dependent process and as such depends critically on the conducting and magnetic properties of the few atomic layers next to the insulating barrier. When two dissimilar oxides are put together, such as the case of manganite/tunnel barrier in all oxide magnetic tunneling junctions (MTJ), electronic and structural reconstruction at the interface, controlled primarily by elastic strain and electrostatic boundary conditions, may modify substantially *3d* orbital filling breaking the $e_g$ orbital degeneracy thus, drastically modifying the magnetic and transport properties.

The study of interfacial effects is a difficult task because in LSMO/tunnel-barrier structures interfaces are buried below several nm of the top electrode. Consequently most approaches have consisted on investigating the film/substrate bottom interface in ultra-thin films. In these cases it is expected that the strain imposed by the substrate will play a dominant role. The substrate constrains the in-plane film cell parameters and imposes its in-plane symmetry operations to the film. The latter, being much weaker than the former, will affect the few first cells close to the film/substrate interface and relax[5,6]. In fact, it has been shown both theoretically[7] and experimentally[8,9,10] that tensile stress favors the $x^2$-$y^2$ orbital occupancy and thus CE-type antiferromagnetic (AF) ordering while compressive strain favors $3z^2$-$r^2$ occupancy leading to a C-type AF ordering. However, X-ray linear dichroism investigations (XLD) in manganite ultra-thin films show that irrespective to the biaxial strain conditions the interfacial Mn atoms show a preferential occupancy of the Mn $3d$ $3z^2$-$r^2$ $e_g$ orbitals[9,10,11]. The reason for this strain-independent selective orbital occupancy is still to be unveiled.

In this work we address the origin of the so-called dead layer appearing close to the manganite/tunnel barrier interface in MTJ. We characterize the magnetic and transport properties at the interface between LSMO and different capping layers (CL) of interest for spintronic applications ($SrTiO_3$, $LaAlO_3$ (LAO) , $NdGaO_3$ (NGO), MgO and Au) as representative of the manganite/tunnel barrier interface in MTJ. Our data reveals the presence of a 1 nm thick AF/insulating layer at the LSMO/CL interface which is independent of the capping material and very much alike to that found at the free surface of LSMO thin films. Our results indicate that the structural disruption, with its intrinsic inversion symmetry breaking, and in some cases crystal symmetry breaking, at the interfaces lay at the origin of the so-called dead layer.

LSMO samples used in this work were prepared by using radio frequency (RF) magnetron sputtering on top of (001)-oriented $SrTiO_3$ (STO) substrates. Prior to LSMO growth the STO substrates were cleaned in an ultrasonic bath with Milli-Q water and annealed at 1000 °C in air for 2 h to obtain a typical morphology of terraces and step with unit cell height ($\approx$ 0.4 nm), thus selecting mostly a unique atomic termination, likely to be $TiO_2$. The thickness of the samples (~40nm) was determined by using grazing incident x-ray reflectometry. The thickness of the capping layers ($t_c$ ~1.6 ±0.2 nm) was determined by controlling the evaporation time after a careful calibration of the growth rate of each of the different materials used. It was also checked "a posteriori" by using high resolution transmission electron microscopy (HRTEM) obtaining good agreement with the nominal values (see Fig. 1). Further details regarding sample preparation can be found elsewhere[12]. Reciprocal space mapping was performed using a Bruker D8 GADDS system equipped with a 2D

Hi-Star x-ray detector to determine the degree of strain of the films. LSMO films are in-plane fully strained [13], which according to previous results will favor the preferential occupancy of $x^2-y^2$ orbitals[7-10].

Our previous study of these bilayers by means of X-ray absorption spectroscopy (XAS) at the Mn $L_{3,2}$-edges made evident that at the LSMO/barrier layer interface a complex scenario evolves in which the stoichiometric double exchange ferromagnetic and metallic (DE-FM) LSMO phase coexists with other non-ferromagnetic phases (NFM) containing $Mn^{2+}$ or excess of $Mn^{4+}$ species[13] (see Fig. 2a and 2b). Only in the case of the LSMO/LAO interface the XLD spectrum was completely analogous to that of the Mn $L_{3,2}$ spectrum measured for the uncapped LSMO sample and representative of bulk LSMO.

The magnetic characterization of the interface has been done by using X-ray linear dichorism (XLD). The synchrotron experiments were performed at the electron storage ring BESSY II by using the 70 kOe high-field end station located at the UE46-PGM1 beamline. The XLD experiments were done well within the FM phase ($T_c \approx 370$ K) of the perovskite compounds (T=10 K) using incoming horizontal linearly polarized radiation at the Mn $L_{3,2}$-edges. A magnetic field of 20 kOe was applied along the beam propagation direction, saturating the sample magnetization in order to eliminate the FM contribution to the XLD spectra thus leaving alone only AF and anisotropic orbital occupancy contributions. Two spectra were obtained at θ=90° ($\beta^{90}$) and at θ=30° ($\beta^{30}$) so to gain sensitivity to in-plane $x^2-y^2$ and out-of-plane $3z^2-r^2$ oriented $e_g$ Mn $3d$ orbitals, respectively (see inset Fig. 2a). Absorption spectra were obtained by using total electron yield detection mode. The escaping depth of the secondary photoelectrons (2–5 nm) guarantees that the measured spectra are mainly determined by the Mn atoms close to the interfacial region. The XLD is defined as XLD = $\beta^{90} - \beta^0$, being $\beta^0 = 4/3(\beta^{30} - 1/4\, \beta^{90})$, (see ref. 14).

The XLD spectrum (XLD=$\beta^{90}$- $\beta^0$) corresponding to the uncapped sample is depicted in Fig. 2a.

Comparison with previously reported temperature-dependent XLD spectra[11] allows identifying the presence of an AF phase whose magnetic axis is aligned perpendicularly to the sample´s plane. Note that the experimental conditions (see experimental section) guarantee that the XLD signal arises only from AF and anisotropic orbital occupancy contributions. The XLD curve for the LSMO surface is identical to that reported by Aruta et al. in ref. 11 (see Fig. 4b) as indicative of a C-type AF phase which originates from a preferential Mn 3$d$ 3$z^2$-$r^2$ $e_g$ orbital occupancy, pointing out that at the topmost LSMO layers close to the free surface the $e_g$ orbital degeneracy is broken. Our data go a step further and reveals that, irrespective to the capping layer material, a breaking of the $e_g$ orbital degeneracy takes place at the LSMO/capping layer interface. The XLD spectra for the different capping layers are shown in Fig. 3. The closeness in their spectral shape indicates that the AF phase observed in the uncapped film is also present at the interface of LSMO with different capping layers, irrespective to the capping material. Moreover, a quantitative analysis of the XLD spectra demonstrates that this similarity extends not only to the spectral shape but also to its amplitude. The amplitude, defined as 100·$I_{|XLD|}$/$I_{XAS}$, where $I_{|XLD|}$ stands for the integrated intensity of the |XLD| spectrum and $I_{XAS}$ corresponds to that of the absorption spectrum β is similar for all samples. Comparison with our previous results show that the XLD amplitude is neither correlated neither to the NFM phase content (table 1) nor to the interface magnetization reduction deduced from XMCD experiments[13]. However, the XLD amplitude exhibits a clear correlation with the DE-FM phase with the nominal $Mn^{3+}$/$Mn^{4+}$ valence balance (Fig. 4), thus indicating the intrinsic character of this interfacial layer. These results therefore reveal a complex scenario at the manganite's surface and interfaces in which FM, NFM and AF phases coexist. Both NFM and AF phases depress interfacial magnetotransport properties. The NFM phase is characterized by the existence of $Mn^{2+}$ or by an excess of $Mn^{4+}$ atoms and is sensitive to the absence/presence and nature of the capping layer. On the other hand, the AF phase seems to be bounded to the amount of the 2/3-1/3 DE-FM LSMO phase at the surface/interface. This fact, in combination with the surface sensitivity of the TEY detection technique suggests that the AF phase extends homogenously at the interface, i.e. it is indeed an interfacial layer.

To farther characterize this AF phase we have investigated the transport properties across the interface for the samples capped with LAO, STO ,NGO , and MgO by means of an atomic force microscope (AFM) system working in the current sensing (CS) mode (see Ref. 15 for details). As expected, I(V) characteristic curves across the LSMO/CL interfaces exhibit the typical features of a tunneling conduction process (see Fig. 5). The thickness of the insulating layer has been estimated from a quantitative analysis of the I-V characteristic curves by using the Simmons model in the intermediate voltage range given by the equation[16]:

$$J = \left(\frac{e}{2\pi h t^2}\right)\left\{\left(\varphi_0 - \frac{eV}{2}\right)\exp\left[-(2m)^{1/2}\frac{4\pi t}{h}\left(\varphi_0 - \frac{eV}{2}\right)^{1/2}\right] - \left(\varphi_0 + \frac{eV}{2}\right)\exp\left[-(2m)^{1/2}\frac{4\pi t}{h}\left(\varphi_0 + \frac{eV}{2}\right)^{1/2}\right]\right\}$$

(1)

Being $\varphi_0$ the barrier height and $t$ the insulating barrier thickness. Some examples of these fits are shown in Fig. 5. The obtained values for $\varphi_0$ and $t$ are summarized in Table I. The values obtained for $\varphi_0$ are in good agreement with those previously reported[15,17,18,19]. Our analysis yields values for the insulating barrier thickness, $t$, larger by $\Delta t \sim$ 1 - 2 nm than those corresponding to the nominal thickness of the CL (see Fig. 1). This indicates that the effective insulating barrier thickness is increased by interfacial effects in the LSMO topmost layers. In principle, there are two possible sources for explaining this extra insulating layer at the interface, i.e. the NFM and the AF phases. Comparison of the values obtained for $\Delta t$ with the amount of secondary NFM phases detected with XAS (see table 1) for the different interfaces yields an almost perfect linear correlation ($r^2$=0.99) between $\Delta t$ and the amount of NFM. However the intercept for NFM= 0% turns out to be non-zero (~1 nm). This means that in all cases there is an extra ~1 nm thick interfacial insulating layer which is not related to the NFM phase. We argue that this 1 nm insulating layer is related to the AF layer whose existence at the LSMO surface/interfaces has been demonstrated by XLD measurements. Indeed this idea is corroborated by the results obtained for the LSMO/LAO interface in which no NFM phase has been detected but a $\Delta t$ =1 nm is still detected. These results, therefore allow us to assign an effective thickness of ca. 2 ML (~1nm) for the C-type AF-insulating phase present at the LSMO surface and LSMO/CL interfaces.

The origin of this C-AF insulating phase is controversial. The main forces at work by the interface can be grouped in i) structural strain[20], ii) electronic reconstruction[21], and iii) structural disruption and crystal symmetry breaking at the interface[5,22]. In what follows we will analyze each of this contributions in order to clarify their relevance in determining the $e_g$ orbitals occupancy at the interfacial region and therefore, their magnetic and electronic properties, and we will show that according to our experimental results neither structural strain nor electronic reconstruction can explain the results.

i) Structural strain is very important since the substrate constrains the film in-plane cell parameters. Indeed our samples are in-plane fully strained ($a_{Film} = a_{Subst}$ ~0.3905 nm) with a slightly reduced out-of-plane cell parameter $c$ (~0.3870(3) nm). This elongation of the in-plane cell parameters would favor the preferential occupancy of the $x^2-y^2$ orbitals due to the combined effect of the anisotropic change of the cell parameters and therefore of the hopping amplitudes leading to a CE type-AF phase[5,7,8]. However our XLD data, corresponding mainly to the LSMO/capping layer interfacial region, exhibit clear features indicative of a $3z^2-r^2$ preferential orbital occupancy promoting a C type-AF ordering irrespective to the capping layer material and the structural strain. In fact similar discrepancies showing that, irrespective to the strained state of the LSMO films, the substrate/LSMO interface exhibits a $3z^2-r^2$ preferential orbital occupancy promoting a C-AF ordering, have already been reported in very thin LSMO films[9,10,11]. Consequently it is clear that strain alone cannot explain the observed preferential occupancy of the $3z^2-r^2$ orbitals neither at substrate/LSMO nor at LSMO/surface of LSMO/capping layer interfaces.

ii) The electronic structure of the LSMO film and the capping layers interact at the interface and possible delocalization effects should be taken into consideration. For instance, in the case of LSMO/STO interfaces it is expected that the $e_g$ manganese orbitals delocalize to some extend into the Ti $3d$ empty states. Interaction that is particularly favored in the case of the $3d$ $3z^2-r^2$ orbitals of the Mn and Ti atoms. This delocalization energy tends to favor a Jahn-Teller distortion increasing the occupancy of the $3z^2-r^2$ orbitals of the Mn and therefore the C-AF phase[5]. Although the expected preferential occupancy in that case agrees with our experimental finding, interfacial electronic effects can also safely be rule out based in our experimental results. First, the amplitude of the XLD signal is similar in both capped and uncapped LSMO films pointing to a minor (if any) effect of the capping on the appearance of the C-AF phase. Second, the effect is observed for different CL materials such as, for instance STO, with neutral SrO or $TiO_2$ layers, NGO, with charge unbalanced NdO or $GaO_2$ layers or LAO were the above mentioned hole-doping mechanism does not apply[23].

iii) Finally, it has to be considered that the interface introduces a structural disruption that could also imply the breaking of the crystalline and inversion symmetry of the system which may strongly modify the electronic properties of the LSMO interfacial layers. In fact, it has been theoretically predicted that at the doped manganites surfaces the charge state of the Mn ions is strongly modified leading to charge localization and the preferential occupancy of the $3z^2-r^2$ orbital within the 2 ML closer to the surface[24], i.e. the formation of a thin C-type AF phase at the surface. Our experimental results obtained in the uncapped LSMO film clearly confirm these theoretical predictions. We do observe the existence of an AF phase at the LSMO surface by XLD and this AF layer is also observed in LSMO/capping layer interfaces. In addition transport measurements in the bilayers allow demonstrating its insulating character and estimating its thickness (~ 1 nm) also in excellent agreement with the theoretical predictions[24].

We would like to emphasize the fact that the strength of the detected C-AF phase is independent of the capping layer material and similar in all the cases to that of the LSMO uncapped surface. From these results we conclude that a robust C-AF phase arises at the surface of LSMO films right after growth due to the structural disruption and the breaking of the crystal symmetry and once formed this AF phase is strong enough to persist after capping.

The AF/insulating layer here detected offers a clear explanation of the microscopic origin of the so-called "*dead layer*" that has been often studied in the case of manganite/substrate interfaces, where structural strain is very relevant, but scarcely investigated at the interface which determines the magneto-transport properties in oxide-based MTJ, i.e. the interface with the insulating barrier. Moreover, it might also explain why the improvement of the microstructural quality of interfaces in manganite MTJ although leading to an increase of the temperature range where TMR can be observed does not lead to the observation of a sizable room temperature magnetoresistance.

As we have shown, the capping of the LSMO samples leads in most cases to the appearance of a NFM phase. In addition, due to the structural disruption, we do also detect the existence of an AF/insulating phase which roughly speaking should be considered as concomitant to the interface. The advance in thin film deposition techniques might have yield to obtain almost perfect interfaces from a chemical and strain point of view, thus reducing the presence of the NFM phase. However, the structural disruption and breaking of the crystal symmetry after the growth of the manganite film, and thus the appearance of the concomitant robust C-type AF phase, is unavoidable.

These results allow envisaging the correct strategy for the fabrication of manganite-based tunneling devices. Our results suggest that in order to avoid the degradation of the performances of manganite layers at the interfaces the insulating barrier has to be chosen to avoid the structural disruption, i.e. for example taking advantage of the very rich phase diagram of these materials and modifying the doping rate to obtain an AF insulating phase for the barrier. This could explain the good performance of $La_{0.67}Ca_{0.33}MnO_3$ /$La_{0.3}Sr_{0.7}MnO_3$/ $La_{0.67}Ca_{0.33}MnO_3$ tunneling junctions presenting a TMR values vanishing only 10 K below the $T_C$ of one of the electrodes[26].

In conclusion, the combination of spectroscopic techniques with transport measurements at room temperature, both sensitive to the LSMO interface, highlights a complex scenario at the manganite thin film surfaces and interfaces in which FM, NFM and AF phases coexist. Transport measurements show that the disruption of the DE-FM phase occurs only at the interface where a thin insulating layer of about 2-4 ML is present. This insulating layer is linked to the NFM and AF phases. However, in addition to this, we have detected the existence of a residual 2 ML thick AF-insulating layer whose existence is concomitant to the nominal $Mn^{3+}$/$Mn^{4+}$ DE-FM phase. Its origin lies on the preferential occupancy of the $3d\,z^2\text{-}r^2$ $e_g$ orbitals at the interface, due to the structural disruption and breaking of the crystal symmetry[5,22,24] after the growth of LSMO layer, leading to a C-AF ordering.

This AF phase is robust enough to survive after capping and therefore likely to be present in MTJ interfaces. The presence of this phase modifies the features of the tunneling barrier severely affecting the tunneling conduction process.

*Acknowledgements*

We acknowledge financial support from the Spanish MEC (MAT2009-08024), CONSOLIDER (CSD2007-00041), and FEDER program. The research leading to these results has received funding from the European Community's Seventh Framework Program (FP7/2007-2013) under Grant agreement no. 226716. Z.K. thanks the Spanish MEC for the financial support through the RyC program.

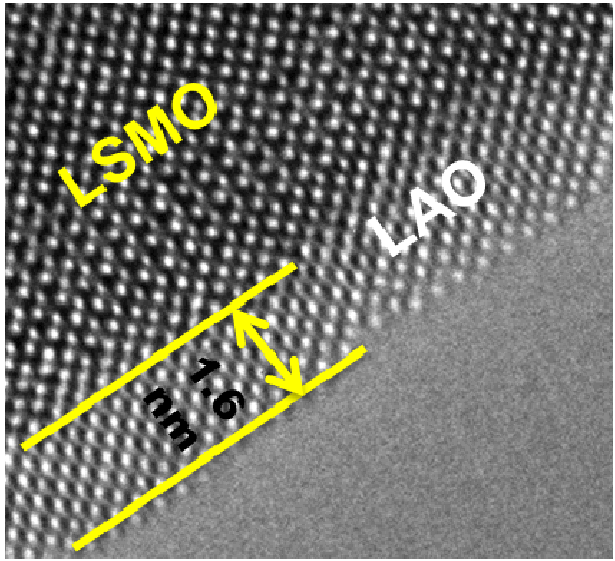

**Figure 1:** HRTEM picture of the LSMO/LAO interface

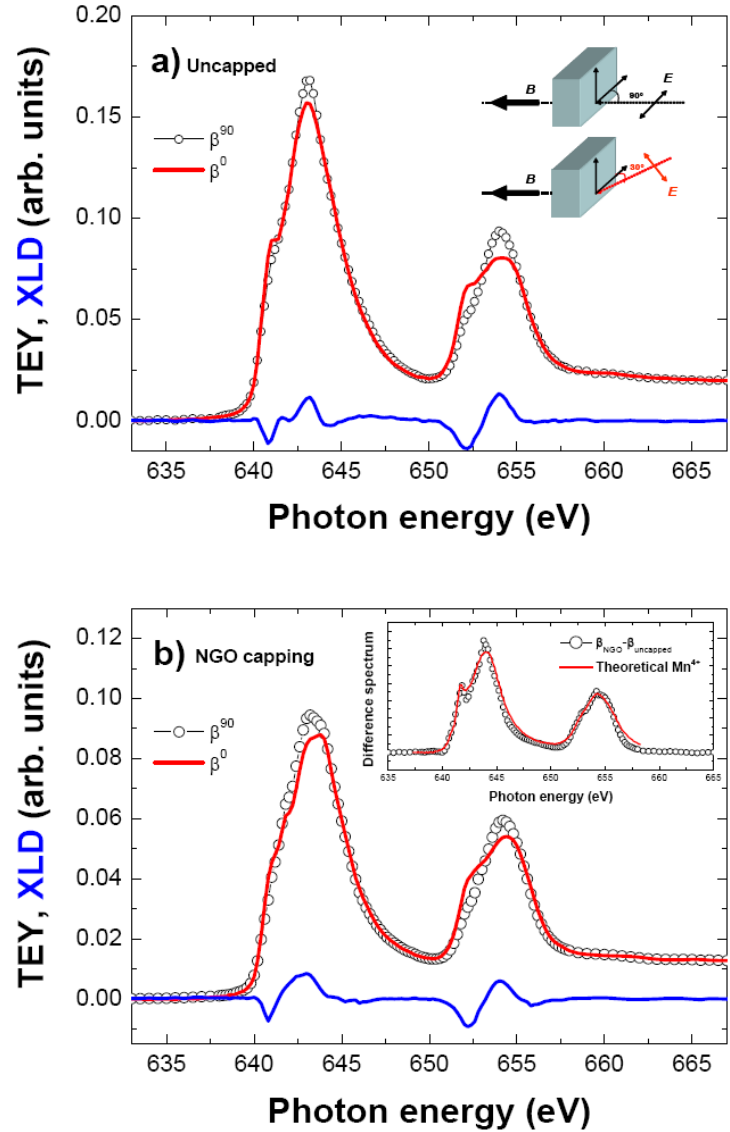

**Figure 2**: *a) and b)* Mn $L_{3,2}$-edge XAS spectra corresponding to the uncapped and NGO-capped LSMO samples, respectively. The electric field vector of the incoming linearly polarized radiation was set parallel ($\beta^{90}$) and almost perpendicular ($\beta^{30}$) to the surface of the sample. Both set of data allowed calculation of the $\beta^0$ spectrum (see text). A magnetic field of 20 kOe was applied parallel to the beam propagation direction saturating the sample magnetization (inset 1a) to remove the FM contribution to the XLD ($\beta^{90}$- $\beta^0$) signal (blue line). *Inset 1b*: Result of subtracting from the LSMO/NGO absorption spectrum ($\beta_{NGO}$) the one obtained for the uncapped LSMO film ($\beta_{uncapped}$) after proper scaling (open dots). The latter, being characterized by showing a bulk-like mixed valence $Mn^{3+}/Mn^{4+}$ spectral shape. The $\beta_{NGO}$ - $\beta_{uncapped}$ difference agrees with the expected shape for a $Mn^{4+}$ (see reference 27)

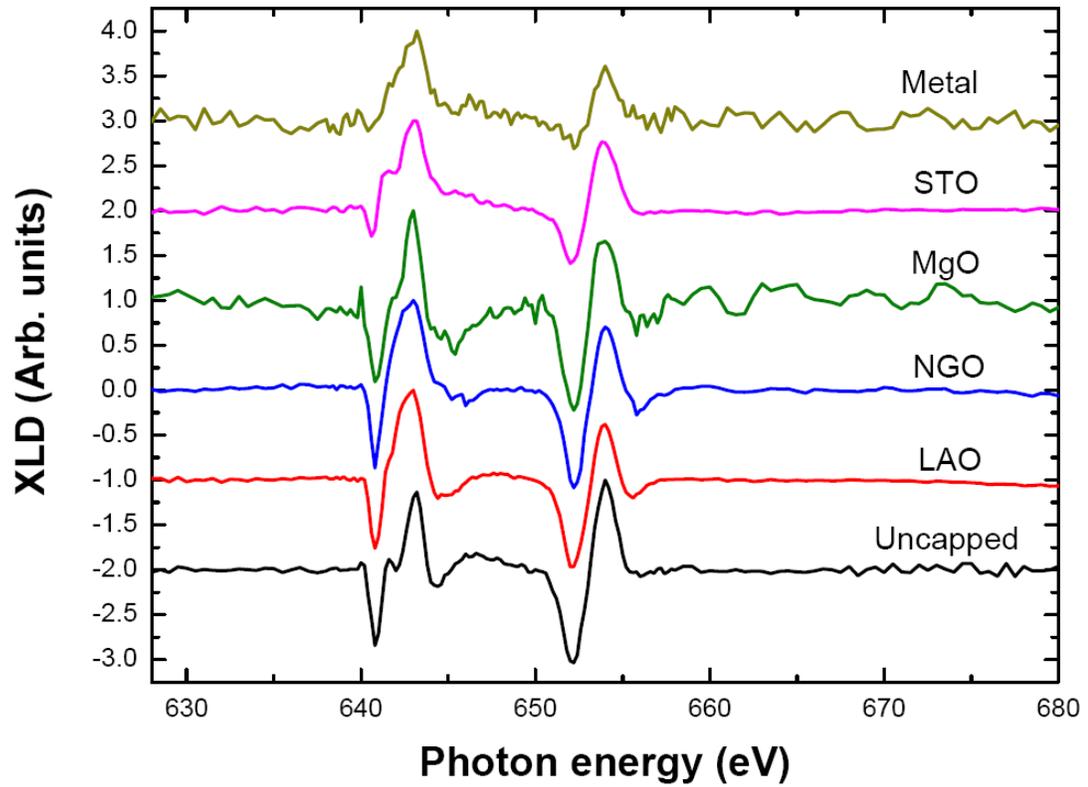

**Figure 3**: XLD spectra obtained at Mn $L_{3,2}$-edge for the various LSMO/capping layer bilayers included in this study. The XLD has been normalized to its maximum value and an offset has been artificially applied for better comparison.

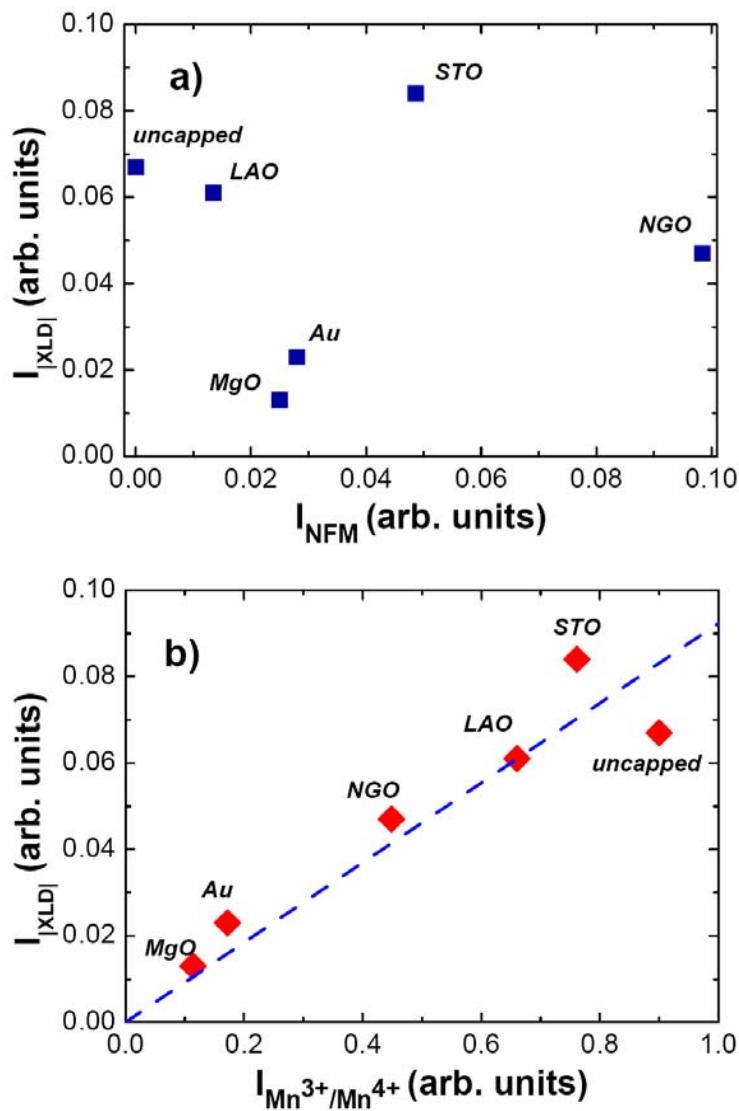

**Figure 4**: Comparison between the integrated intensity of |XLD|, i.e. $I_{|XLD|}$ vs. the integrated intensity of the XAS originating from the $Mn^{2+}$ and $Mn^{4+}$ impurities (top panel) and from a nominal mixed valence $Mn^{3+/4+}$ state, i.e. $I_{Mn^{3+}/Mn^{4+}}$ (bottom panel). A clear linear relation is observed only between $I_{|XLD|}$ and $I_{Mn^{3+}/Mn^{4+}}$. The dashed line shows a fit intercepting the x and y axis at 0.

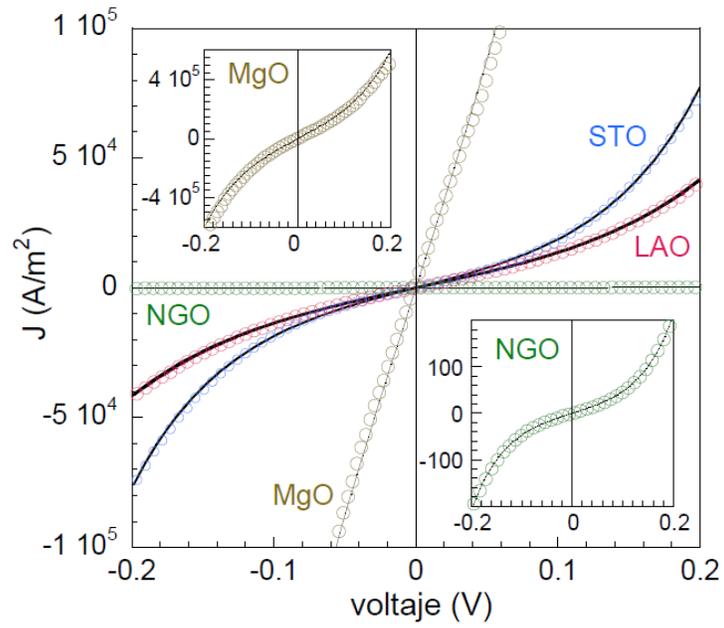

**Figure 5:** Dependence of the current density as a function of the voltage bias, J(V), across the LSMO/capping layer interface for different capping layers. Continuous line corresponds to the fitting by using Eq. [1] in the text. Insets: Detail of the J(V) curve for the LSMO/MgO (upper) and LSMO/NGO (lower) systems.

**Table 1.** Values of the barrier thickness, *t*, and the barrier height, $\varphi_0$, for the different capping layers obtained by using Eq.[1] in the low bias voltage regime. Δt correspond to the difference between t and the thickness of the capping layer (~1.6nm). NFM corresponds to the fraction of phase with Mn oxidation state differing from the nominal $Mn^{3+/4+}$ mixed valence state. The relatively low value of *Δt* in the case of MgO deserves further investigation and might be related to the active role of the MgO barrier and its spin filtering effect. It is also worth mentioning here that for MgO capping J(V) curves are slightly asymmetric and fittings by using Eq.[1] are worse than in the rest of the cases. This might also affect the evaluation of *Δt* for MgO capping.

| Capping Layer | φ0 (eV) | t (nm) | Δt ~ t-tc (nm) | NFM (%) |
|---|---|---|---|---|
| LAO | 0.45 | 2.6 | 1 | 0% |
| STO | 0.32 | 3 | 1.4 | 6% |
| NGO | 0.41 | 3.6 | 2 | 18% |
| MgO | 0.38 | 2.5 | 1 | 18% |